\documentstyle[12pt]{article}

\def\nue{ \nu_e  }
\def\numu{ \nu_\mu  }
\def\nutau{ \nu_\tau }
\def\dm2{ \Delta m^2 }
\def\s22t{ \sin^22\theta }
\def\Tc{T_c}
\def\beq{ \begin{equation} }
\def\eeq{ \end{equation} }
\def\beqa{\begin{eqnarray} }
\def\eeqa{\end{eqnarray} }
\def\beqa*{ \begin{eqnarray*} }
\def\eeqa*{ \end{eqnarray*} }

\def\PR{ {\it Phys. Rev.} }
\def\PRL{ {\it Phys. Rev. Lett. } }
\def\PL{ {\it Phys. Lett. } }
\def\RMP{ {\it Rev. Mod. Phys.} }
\def\NP{ {\it Nucl. Phys. } }
\def\NC{ {\it Nuo. Cim.} }
\def\etal{{\it et al.}}
\begin{document}
\vspace*{-10ex}
\vspace{1.0ex}

\renewcommand{\thefootnote}{\fnsymbol{footnote}}
\setcounter{footnote}{1}

\begin{center}
   {\large
         Implications of Combined Solar Neutrino Observations   \\
              and Their Theoretical Uncertainties
     } \\

    \vspace{2.6ex}

   S.\ A.\ Bludman,
   N.\ Hata,
   D.\ C.\ Kennedy\footnote{
                                Present address:
                                Fermi National Accelerator Laboratory,
                                P.O. Box 500 MS106, Batavia, IL 60510  },
   and P.\ G.\ Langacker \\

    \vspace{1.5ex}

{\it
   University of Pennsylvania, Philadelphia, PA 19104    \\
}
                    July 2, 1992, UPR-0516T

\vspace{2.6ex}
\noindent   {\bf Abstract }

\end{center}

Constraints on the core temperature ($T_c$) of the Sun and on
neutrino-oscillation parameters are obtained from the existing
solar neutrino data, including the recent GALLEX and Kamiokande III
results.
(1)  A purely astrophysical solution to the solar neutrino problem
is strongly disfavored
by the data: the best fit in a cooler Sun model requires an
8\% reduction in $T_c$, but the $\chi^2$ test rejects this
hypothesis at
99.99\% C.L.  This suggests neutrino physics (mass, mixings) that goes
beyond the standard electroweak theory.
(2) Assuming the Standard Solar Model (SSM) and matter-enhanced
neutrino oscillations, the MSW
parameters are constrained  to two small regions: non-adiabatic
oscillations with
$\dm2 = (0.3 -  1.2) \times 10^{-5} \ \mbox{eV}^2, \,
\s22t = (0.4 - 1.5)  \times 10^{-2}$,
or large mixing-angle oscillations with
$ \dm2 = (0.3 - 4) \times 10^{-5} \  \mbox{eV}^2, \,
\s22t = 0.5 - 0.9 $.  The non-adiabatic solution gives a
considerably better fit.
For $\nue$ oscillations into sterile neutrinos, the allowed region (90\%)
is constrained to non-adiabatic oscillations.  As long as the SSM is
assumed, the neutrino mixing angles are at least four times larger,
or considerably smaller, than the corresponding quark mixing angles.
(3) Allowing both MSW oscillations and a non-standard core temperature,
a) the experiments determine the core temperature at the 5\% level:
$T_c = 1.03^{+0.03}_{-0.05}$ (90\% C.L.) relative to the SSM,
which is consistent with the SSM prediction.
b) When $T_c$ is used as a {\it free} parameter, the allowed MSW region
is broadened: a cooler Sun ($\Tc = 0.95$) allows $\dm2, \s22t$
implied by the supersymmetric SO(10) grand
unified theory (GUT), while a  warmer Sun ($T_c=1.05$) extends the
allowed parameter space into values suggested by
intermediate-scale SO(10) GUTs, for which the $\nutau$ may be
cosmologically relevant.  Superstring-inspired models are
consistent with all solutions.
(4) From the narrowed parameter space, we predict the neutrino spectral
shape which should be observed in the Sudbury Neutrino Observatory (SNO).
Expected rates for SNO, Super-Kamiokande, and BOREXINO are also discussed.
Throughout the calculation we use the Bahcall-Pinsonneault SSM (1992) with
helium diffusion, and include
nuclear and astrophysical uncertainties in a simplified, but physically
transparent way.


\setlength{\textheight}{21.0cm}
\setlength{\topmargin}{-0.9cm}
\newpage
\vspace{4.0ex}

\noindent
I.  INTRODUCTION: EXPERIMENTAL STATUS AND SSM IMPROVEMENTS
\vspace{2.0ex}

Since solar neutrinos were first detected two decades ago,
the observed neutrino flux has always been a factor of 1.5 to 4 times
less than that predicted
by the Standard Solar Model (SSM).  When both experimental and SSM
uncertainties
are included, the Homestake chlorine (Cl) experiment, Kamiokande, and GALLEX
rates respectively are approximately
$6\sigma$, $3\sigma$, and $2\sigma$ below the SSM predictions.
Although the most recent GALLEX deficit is only 35\%, significant
differences from the SSM predictions persist.   In this paper, we consider
both astrophysical solutions (a cooler Sun model)
and particle-physics solutions (MSW)  to the solar neutrino problem.
We show that no reasonable change in the SSM can
reconcile the  quoted Homestake, Kamiokande, and
GALLEX results.
On the other hand the MSW effect, which assumes
neutrino mass differences $\dm2$ and mixing $\s22t$,
accommodates all data:
it does not require discarding  any of the experiments,
nor stretching  the SSM beyond its uncertainties.
Once MSW is admitted, the data determine $\Tc$ at the 5\% level, yielding
a  value consistent with the SSM.  We also discuss possible MSW parameters
when a non-standard core temperature is assumed.

There are four measurements of the solar neutrinos so far.  The
chlorine experiment at Homestake \cite{Homestake}
 is mainly sensitive to $^8$B and $^7$Be
neutrinos,  and reports\footnote{
   We ignore the possibility of time dependence in the Homestake data.}
an observed
rate 2.1$\pm$0.3 SNU (0.28$\pm$0.04 of the SSM), while the SSM
prediction
is 8.0$\pm$1.0 SNU \cite{BP}. (Quoted errors are all 1 $\sigma$ in this
paper.)  A direct counting measurement by the Kamiokande Collaboration
 observes
\v{C}erenkov light from recoil electrons scattered by $^8$B neutrinos.
The combined result of Kamiokande II \cite{Kamiokande} and 220 days
of Kamiokande III \cite{Kamiokande III}
is 0.49$\pm$0.08 of the central value of the SSM; there is
an additional 14\% uncertainty in the SSM prediction.
A unique opportunity to observe
low energy pp neutrinos, which come from the main reaction
responsible for the energy generation in
the Sun, is provided by the SAGE and GALLEX
gallium experiments.
The GALLEX result is 83$\pm$21 SNU (0.63$\pm$0.16 of the SSM)
\cite{GALLEX},
again well below the SSM value 132$^{+7}_{-6}$ SNU.  SAGE has published
a lower value 20$\pm$38 SNU based on their first 5 extractions
\cite{SAGE}.
The later SAGE runs clearly suggest a higher rate, but no cumulative rate
has been published.
We will therefore exclude the SAGE results from our analysis.

The SSM itself has undergone several refinements
over the years.  There are presently at least four SSMs
\cite{BP,BU,SBP,TC},
which, for the same physics input, agree with each other within 1\%
\cite{BP}
and agree with the sound speed calculated in p-mode helioseismology
within 2\% \cite{TC-Granada}.
The latest published solar model by Bahcall and Pinsonneault improves on
earlier calculations by using the most recent OPAL calculated opacities,
meteoritic iron abundances, and updated nuclear-reaction
cross sections \cite{BP}.
The new model includes the effects of helium
diffusion so that the observed $^3$He surface abundance is obtained.
Their prediction for the convective zone boundary is in striking agreement
with helioseismology data \cite{helioseismology}.
The surface abundances of the $^7$Li and $^7$Be are still overestimated,
but the surface abundances have negligible effect on
the neutrino production in the core.

\vspace{4.0ex}
\noindent
II.  COOLER SUN WILL NOT EXPLAIN THE OBSERVED NEUTRINO FLUXES
\vspace{2.0ex}

The production of high energy
$^8$B neutrinos and  intermediate energy $^7$Be neutrinos is
directly proportional to the
$^3$He$(\alpha,\gamma)^7$Be and $^7$Be(p,$\gamma)^8$B
nuclear cross sections and depends very sensitively on the solar
temperature in the innermost 5\% of the Sun's radius.  This inner core
temperature is not probed by existing p-wave
helioseismological observations,
but is determined by the radiative opacities throughout the Sun, which
are believed to be calculated to within a few percent \cite{BP}.

Many nuclear and astrophysical explanations have been proposed
to explain the solar neutrino deficit.
One possibility is to change input
parameters, such as lowering the $^7$Be(p,$\gamma)^8$B cross section
or reducing the opacity.  Another is to invoke mechanisms that are
not included in the SSM, such as a large core-magnetic field,
a rapidly rotating core, or hypothetical WIMPs which carry away
energy from the core.  The net effect is generally a lowering of the
core temperature and thus a reduction of the nuclear burning taking
place there.

It is therefore reasonable to examine the core temperature as a
diagnostic of a whole class of astrophysical explanations of the
neutrino deficit, although the temperature profile is the result, not
an input, of a solar model  \cite{cool Sun}.
We choose the
central temperature ($T_c$) as a phenomenological parameter representing
different conceivable solar models.  The approximate
correlation of the neutrino fluxes with $\Tc$ was obtained by
Bahcall and Ulrich \cite{BU} by examining
1000 self-consistent SSMs with randomly distributed
input parameters, and is given as simple power laws \cite{BU}:
\beq
     \phi(\hbox{pp}) \sim T_c^{-1.2},  \hspace{5mm}
     \phi(^7\hbox{Be}) \sim T_c^8   ,  \hspace{5mm}
     \phi(^8\hbox{B }) \sim T_c^{18} .
\eeq
For each experiment we therefore parameterize
rates relative to the SSM as functions of $\Tc$:
\beqa*
    R_{Cl} &=& (1 \pm 0.033)
                  [ 0.775\times (1 \pm 0.100) \times T_c^{18}
                   +0.150 \times(1 \pm 0.036)\times T_c^8    \\
           &&      + \hbox{small terms} ]                   \\
    R_{Kam}&=&  (1 \pm 0.100) \times \Tc^{18}               \\
    R_{Ga} &=&  (1 \pm 0.04)[
                   0.538\times(1 \pm 0.002)\times\Tc ^{-1.2}
                 + 0.271\times(1 \pm 0.036) \times \Tc ^8    \\
           &&    + 0.105\times(1 \pm 0.100) \times \Tc^{18}
                 + \hbox{small terms} ],
\eeqa*
where $\Tc$ is the central temperature relative to the SSM
($\Tc = 1 \equiv 15.67 \times 10^6$ K).
The Bahcall-Pinsonneault solar model including
diffusion \cite{BP}
is used throughout the paper, unless otherwise mentioned.
The uncertainty in the overall
factors for Cl and Ga is due to the detector reaction
cross sections.
The uncertainties in each neutrino flux include only nuclear physics
uncertainties in production cross sections;
astrophysical uncertainties are absorbed into a variable $\Tc$.
These flux uncertainties
are properly correlated for the three experiments.

The $\Tc$ dependence of the Kamiokande, Cl, and Ga detectors are
each shown in Fig.~1, and the best fits for various combinations
of the data are
summarized in Table~I.  If each experiment were fit alone, Kamiokande
and Homestake require a reduction of
$\Tc$ by 4$\pm$1\% and 10$\pm$1\%
respectively.  For GALLEX alone, $\Tc$ must be
reduced by 14$\pm$4\%, but still
does not completely fit the data, because the negative exponent of $\Tc$
in the pp flux works against the reduction of the total rate.
In order to match the observed luminosity, the pp chain has to
compensate the energy loss due to reductions of the other reactions.
As a result the $\Tc$ reduction
fails to fit the central value of the GALLEX rate,
although it is compatible with the upper end of the range.
The three separate $\Tc$ fits are respectively
3$\sigma$, 6$\sigma$, and 3$\sigma$ below the SSM prediction
$\Tc = 1 \pm \Delta \Tc$, where the SSM uncertainty
$\Delta \Tc = 0.0057$ is estimated from the 1000 SSM calculations
in Ref.~\cite{Tc error}.

The combined observations cannot be fit by any single $\Tc$.
The larger Kamiokande rate relative
to the Cl rate especially contradicts the $\Tc$ dependence shown in
Fig.~1.  The simultaneous fit of Kamiokande and Cl yields
$\Tc$ = 0.92$\pm$0.01, but the $\chi^2$ value is so large
($\chi^2$=13.78) as to
exclude the fit at the $>$99.99\% C.L.  The combined Kamiokande and GALLEX
results yield a marginally consistent $\Tc$: the best fit is
$\Tc$ = 0.96$\pm$0.01 with $\chi^2 = 2.64$ (89.3\% C.L.).  When all three
experiments are fit simultaneously,
$\Tc$ = 0.92$\pm$0.01 but the $\chi^2$ test
rejects the cooler Sun hypothesis at 99.99\% C.L.  This strong rejection
of the cooler Sun is driven mainly by the contradiction between
Kamiokande and Homestake and, secondarily, the low GALLEX result.

In our fit, $\Tc$ is
allowed to vary in a range wider than that for which the power
laws (Eqn.~(1)) were derived.
We assume that, even for large changes of $\Tc$, the power laws yield
a reasonable approximation to the core-temperature dependence
of the neutrino fluxes.
It should be noted that our conclusions do not depend on
a specific choice of the exponents given in Eqn.~(1).
Provided that
the $^7$Be flux is less temperature dependent than the $^8$B flux, the
Cl rate is expected to be larger than the Kamiokande rate,
contradicting the data \cite{BKL}.
Even if both flux components had
the same temperature exponent (=18), we find that
$\Tc = 0.93 \pm 0.01$ and $\chi^2 = 10.21$ for 1 degree of freedom:
the $\chi^2$ test excludes the fit at 99.91\% C.L.,
with similar conclusions for other exponents.
Based on these observations
one can conclude that the cooler Sun model, which is a nearly universal
feature of astrophysical solutions to the solar neutrino problem, is
strongly disfavored by the data: the central value of
the GALLEX result and the combined result of Homestake and Kamiokande
are each incompatible with the cooler Sun hypothesis.

\vspace{4.0ex}
\noindent
III. MSW FIT TO THE COMBINED OBSERVATIONS
\vspace{2.0ex}

While  modifications of the solar model cannot accommodate
the data, an attractive solution is proposed from particle physics.
Matter-enhanced
neutrino oscillation (MSW effect), first proposed by
Wolfenstein, then applied to solar neutrinos by Mikheyev and Smirnov,
offers a natural explanation of the observed solar neutrino deficit
without requiring any ad-hoc mechanisms \cite{MSW}.
This MSW mechanism assumes new
properties of neutrinos, mass and mixings, to convert electron-neutrinos
to other species when the neutrinos
propagate through the Sun, making possible a large reduction
of the $\nue$-counting rate.
The MSW assumption of neutrino mass and mixing
is a natural extension of the Standard Model,
and requires no ad-hoc
features such as a large magnetic moment.  Unlike vacuum
oscillations, it does not require fine-tuning.  If the
MSW oscillation takes place in the Sun, the determination of the neutrino
mass and
mixing will provide a clue to grand unified theories, which naturally
lead to parameters in the relevant region \cite{BKL,CL}.

The MSW effect depends on two parameters: the mass-squared
difference $\dm2$ and the vacuum mixing angle $\theta$ between $\nue$ and
another neutrino species into which it converts.  The conversion
occurs in a wide parameter space that covers four orders of magnitude
both in $\dm2$ and $\s22t$: a triangle-shaped region in the
$\s22t$ vs. $\dm2 /E$ plane, surrounded by
$\dm2 /E \leq 2\times 10^{-5}$ (eV$^2$/MeV) and
$\s22t \cdot \dm2 /E \geq 10^{-9}$ (eV$^2$/MeV),
where $E$ is neutrino energy.  Typical survival-probability contours
are shown in the
$\dm2$ vs. $\s22t$ plane (MSW diagram) after
integrations over the neutrino production site and the neutrino energy,
including the detector cross sections \cite{Haxton}.

The MSW diagrams show
three physically-distinct regions: the adiabatic region
(the horizontal upper arm of the triangle),
the non-adiabatic region (the diagonal arm), and the large-mixing
region (the right, vertical arm).
For the adiabatic solution, the MSW resonance takes place in the core
of the Sun where the neutrinos are produced; the density is high enough
for the higher-energy neutrinos to resonate and be
depleted while the lower-energy ones survive.
For the non-adiabatic solution, the higher-energy
neutrinos survive more because of non-adiabatic (Landau-Zener) jumping.
In the large-mixing region, which connects smoothly to vacuum
oscillation,
the neutrino spectrum is equally reduced
over the whole spectrum. In the middle of the isosnu triangle, almost 100\%
conversion of $\nue$ occurs.

This flexibility
makes the MSW effect phenomenologically robust.   It can
preferentially suppress the high energy ($^8$B)
neutrinos, or the low energy ($^7$Be and pp) neutrinos more.
It can deplete the lower energy part of the
$^8$B and $^7$Be spectrum while keeping the pp flux,
as suggested by the experiments (Fig.~14).

In the MSW calculations, we use
the Parke formula \cite{Parke}  instead of solving the Schr\"{o}dinger
equation for the oscillations
numerically.  The jump probability given by the
exponential Landau-Zener formula
$ P_j = e^{-\chi} $ is used \cite{Pizzochero},  where
$
\chi = \pi h \sin^2\theta \dm2 /E,
$
$E$ is the neutrino energy, and $h = (-d\log n_e/dr)^{-1}$
is the electron-density scale height
evaluated at the resonance point.  This formula agrees with
the exact solution for large mixing angles and with the linear Landau-Zener
approximation \cite{Landau-Zener} for the small mixing region.
The spatial distribution
of neutrino production in the Sun is taken from Bahcall and Pinsonneault
\cite{BP}.  The
detector cross sections are taken from Bahcall and Ulrich \cite{BU}.
For Kamiokande,
the $\numu$ (or $\nutau$) contribution for flavor oscillations,
the energy threshold, the energy resolution, and the trigger efficiency
are all properly included \cite{KII-PRD}.

In fitting the data, theoretical uncertainties of the SSM are treated
with care, using a simple and transparent parameterization.
For each flux component the nuclear cross section uncertainties of
every reaction are added quadratically to the detector cross section
uncertainties and to the astrophysical (non-nuclear) uncertainties.
The latter is represented by the
uncertainty in the central temperature $\Delta \Tc$ times the exponent
defined in Eq (1).
The theoretical uncertainty  $\Delta \Tc=0.0057$ is chosen to yield
 flux uncertainties consistent with those given in
Ref.~\cite{BP}, and to be consistent
with estimates from the 1000 SSM Monte-Carlo calculations of
Bahcall and Ulrich \cite{Tc error}.
The correlations of the uncertainties among the experiments and flux
components are properly
taken into account.
Our calculations were compared with other studies which utilize
1000 Monte-Carlo SSMs \cite{BH,SSB}; the agreement is excellent.

Effects of the astrophysical uncertainties on the MSW effect were
also examined.  Both the uncertainties from the neutrino production
profile, which affects the matter mixing angle at the neutrino
production, and
from the electron-density scale height, which enters in the jump
probability, were found to be small.

Survival-probability contours and 90\%-C.L. allowed regions are shown
for each experiment in Fig.~2,~3\,(a),~4 and~5.
The Homestake allowed region (Fig.~2) does not
precisely trace the iso-probability contour because of the difference in
$^7$Be and  $^8$B theoretical
uncertainties.  Fig.~3(a) and 3\,(b) show the
Kamiokande result for flavor oscillations and oscillations to sterile
neutrinos, respectively.

The combined result (90\% C.L.) of Homestake, Kamiokande II+III, and
GALLEX is displayed in Fig.~6\,(a).
The confidence-level region is defined  from $\chi^2$ values that satisfy
$
  \chi^2(\s22t ,\dm2 ) = \chi^2_{min} + 4.6,
$
which is valid in the approximation
that the allowed regions are `ellipses' on the
$\log\s22t-\log\dm2$
plane.  Including the GALLEX observations, the allowed MSW parameters
are either
$\dm2 = (0.3 -  1.2) \times 10^{-5} \ \mbox{eV}^2, \,
\s22t = (0.4 - 1.5)  \times 10^{-2}$
(non-adiabatic solution), or
$ \dm2 = (0.3 - 4) \times 10^{-5} \  \mbox{eV}^2, \,
\s22t = 0.5 - 0.9 $
(large-mixing solution).
The best fits of $\dm2$ and $\s22t$ along with $\chi^2$ value for each
region are
listed in the second and third columns of
Table~II.  The experiments prefer the
non-adiabatic solution to the large-angle solution.
The non-adiabatic MSW solution yields a good fit ($\chi^2$=0.56); but
in the large-mixing region, the $\chi^2$ value is large
($\chi^2 = 3.52$): it is
allowed at the 90\% C.L. by the definition above, but for 1 degree of freedom
(= 3 experiments $-$ 2 parameters) this region is excluded at 95\% C.L.
The allowed regions at the 68, 90, and 95\% C.L. are shown
in Fig.~6\,(b):
there is no parameter allowed in the large-angle region at 68\% C.L.
The combined fit without theoretical
uncertainties from the SSM and detector cross
sections is displayed  in Fig.~6\,(c).  Comparison with Fig.~6\,(a)
shows the noticeable effect of SSM uncertainties.
Our Fig.~6\,(a) practically agrees with that obtained by the GALLEX group
\cite{GALLEX}, who included
the day-night effect and $\nue$ regeneration in the Earth, which we
have neglected.

Allowed regions for various combinations
of any two experiments are shown in Fig.~7 (Kamiokande and Homestake),
Fig.~8 (Kamiokande and GALLEX), and Fig.~9 (Homestake and GALLEX).

We have also examined possibilities of oscillations to a
sterile neutrino
\cite{sterile,17-keV}.
If $\nue$  oscillates into a sterile neutrino instead of $\numu$ or
$\nu_{\tau}$,
the term $n_e - n_n/2$ enters the MSW equation in place of
$n_e$ .
(Here $n_e$ and $n_n$ are the local electron and neutron densities
in the Sun \cite{sterile}.) Also, for the Kamiokande detector there is no
neutral current
contribution from the converted neutrinos.  The result for $\nue$
oscillations into sterile neutrinos is displayed in Fig.~3\,(b) for
Kamiokande, and in Fig.~10 and Table~III
for the combined fit.  (There is no
significant change for the Homestake and GALLEX experiments.)
Because of the constraints by the Kamiokande observations,
there is no solution in the large-angle region at 90\% C.L.,
and the allowed parameters are limited in the non-adiabatic region.
Even in the non-adiabatic region, the
best fit yields $\chi^2 = 3.17$ and is excluded at 92\% C.L. for 1
degree of freedom.  This is because of the smaller
Homestake rate relative to  Kamiokande:
the absence of the neutral current events in Kamiokande requires
a larger
electron survival probability than for the flavor-oscillation case,
and therefore widens the discrepancy between the two experiments.

The precise determination of MSW parameters will allow us to draw
some theoretical conclusions in Sec.~V and to make predictions
for next-generation neutrino experiments in Sec.~VI.

\vspace{4.0ex}
\noindent
IV.   SIMULTANEOUS FIT OF MSW AND CORE TEMPERATURE
\vspace{2.0ex}

We have also studied the possibilities of having both MSW oscillations
{\em and} a non-standard solar model by allowing $\Tc$  to be a
completely free parameter.  (We use the same nuclear and detector
uncertainties as in the SSM case.)
The data are fit simultaneously to three parameters
 $\dm2$, $\s22t$, and $\Tc$.  The $\chi^2$ plot is displayed
as a function of $\Tc$  (Fig.~11) , where
$\chi^2$ is minimized for each $\Tc$
with respect to $\dm2$ and $\s22t$ in the
allowed region in the MSW diagram.
By the $\chi^2$ fit, the data determine the core temperature at
the 5\% level.  The best fits are
$ \Tc = 1.03^{+0.03}_{-0.05} $
(90\% C.L.) in the non-adiabatic region and
$ \Tc = 1.05^{+0.01}_{-0.07} $
in the large-mixing region, in good agreement with the SSM prediction
$\Tc = 1 \pm 0.0057$.
The consistency between the data and the SSM is encouraging.  Moreover,
even allowing the MSW conversion and
the other uncertainties, the observations determine the
core temperature to within 5\% \cite{BB}.

The allowed region for the 3-parameter fit is shown in  Fig.~12.
For each $\dm2$ and $\s22t$, the $\chi^2$ is minimized with respect to
$\Tc$, and
$
  \chi^2(\s22t ,\dm2 ) = \chi^2_{min} + 4.6,
$
determines the 90\% C.L. allowed region, where $\chi^2_{min}$ is
a minimum with respect to all 3 parameters.
By allowing  $\Tc$ to be a free parameter, the allowed region
is widened, now stretching from $\s22t =  4\times 10^{-4} \sim 1$ and
$ \dm2 = (0.2 -  4 ) \times 10^{-5}eV^2$.
The best fit parameters are shown in Tables~II and III.
The $\Tc$ dependence of
the region is seen in Fig.~13\,(a) and Fig.~13\,(b), which shows
the 90\% C.L. contours when $\Tc$ is fixed at 1.05 and 0.95 respectively.
The higher
temperature Sun allows a region between the two islands allowed in the
SSM case (Fig.~6\,(a)),
while the cooler Sun pushes the parameter-space outward.

\vspace{4.0ex}
\noindent
V.  THEORETICAL IMPLICATION OF FITTED MSW PARAMETERS
\vspace{2.0ex}

The best fit MSW parameters from Fig.~6\,(a) and Fig.~11 are summarized
in Table~II for the SSM ($\Tc = 1$) and for non-SSM in which $\Tc$ is
an adjustable parameter.  The results for sterile-neutrino oscillations
are listed in Table~III.
 The $\dm2$ range is consistent with the
general expectations of grand unified theories \cite{BKL} or
string-inspired models \cite{CL}, but the mixing angles are not in
agreement with the expectation $\theta_{lepton} \sim \theta_{CKM}$
of the simplest GUTs \cite{BKL}.
If we accept the SSM, the observed neutrino mixing are
at least four times larger (or considerably smaller)
than the quark mixings,
$\s22t = 0.18$ and $< 2\times 10^{-3}$ for $u-c$ and $u-t$ quarks,
respectively.
If we allow a warm Sun, then Cabibbo mixing with
$\dm2 \simeq ( 0.8 - 2) \times 10^{-5}$ eV$^2$ is possible;
this suggests
$m_{\numu} \simeq (3 - 4) \times 10^{-3}$ eV and an SO(10) GUT
with intermediate-scale symmetry breaking.  If we allow a cool Sun, then
$\dm2 \simeq ( 0.5 - 1.5) \times 10^{-5}$ eV$^2$
is possible, suggesting
$m_{\nutau} \simeq ( 2 - 4) \times 10^{-3}$ eV
and a SUSY SO(10) GUT (with large Higgs representations so that
the seesaw scale is close to the unification scale).

GUT predictions for neutrino masses are much less robust than for mixing
angles and mass ratios.  We therefore regard the seesaw
model \cite{seesaw} only as a crude guide
to neutrino masses.  In the SUSY GUT case, all neutrino masses are
cosmologically and astrophysically insignificant.  In the non-SUSY
SO(10) GUT, the seesaw model suggests cosmologically interesting
$\nutau$ masses.  Since $m_t/m_c \sim 100$ and
$m_\tau/m_\mu = 17$ we have
$m_{\nutau}  \simeq 0.4 - 0.8 \;  (40 - 80)$ eV
for a linear (quadratic) seesaw model with up-quark masses and
$m_{\nutau} \simeq 0.07 - 0.14 \; (1 - 2)$ eV
for a linear (quadratic) seesaw model with charged lepton masses.
In carrying out these extrapolations, we have taken
$m_{\numu} \simeq (2 - 4) \times 10^{-3}$ eV
and, for the up-quark case,
have included a factor two to renormalize mass ratio from from GUT to low
energy scales \cite{BKL}. (This factor increases beyond two non-linearly
for large value of the top quark mass because of the Higgs corrections
to the top mass \cite{DCK}.)
We see that only a intermediate scale quadratic seesaw mechanism
can give cosmologically significant $\nutau$ mass \cite{BKL}.
String-inspired models can generate an intermediate seesaw scale via
effective non-renormalizable operators \cite{CL}.  There are no clear
mixing angle predictions in such models,
but for $\nue \rightarrow \numu$ oscillations
consistent with either MSW solution, the $\nutau$ may again be
cosmologically significant.

\vspace{4.0ex}
\noindent
VI.  PROSPECTS FOR FUTURE EXPERIMENTS
\vspace{2.0ex}

One can predict the results of future solar neutrino
observations from the parameters of the combined fit (Fig.~6\,(a)).
The $\nue$ survival probability is shown as a function of energy
in Fig.~14 for each of the  allowed regions.
The predicted observed rates for the high-energy $^8$B
$\nue-e$ scattering (SNO \cite{SNO}
and Super-Kamiokande \cite{Super-K}),  the $\nue-d$
reaction
(SNO), and the $\nue-e$ scattering from $^7$Be neutrinos
(BOREXINO \cite{Borexino}) are listed in Table~IV \cite{Rosen}.
The detector cross sections
as well as the proposed energy resolution of the
detectors are included in the calculation.

The measurement of the charged current reaction
$\nue+d \rightarrow e+p+p$, which is planned for the first-year
operation of SNO, will be  a clear diagnostic of the MSW;
the distortion of the energy
spectrum is characteristic of most particle physics solutions of the
solar neutrino problem and cannot be caused by
any astrophysical effects operative in the Sun \cite{Bahcall-spectrum}.
Figure~15 shows the predicted energy spectra for both
the non-adiabatic and the large-mixing solution, along with estimated
statistical errors equivalent to a 2-year operation (6000 total events).
The non-adiabatic spectrum is very similar to the predicted spectrum
for sterile neutrinos.
If there is a non-adiabatic MSW effect, as suggested by the
best fit (Table~II),  the spectral distortion will
(a) confirm the MSW effect, and (b) discriminate between the two
presently allowed solutions.  Of course, the neutral to charged
current ratio would also establish MSW oscillations into $\numu$ or
$\nutau$ (but not into a sterile $\nu$).

The measurements of the neutral current events by SNO and BOREXINO,
and their ratio relative to the charged current events
would provide definite evidence for MSW flavor oscillations.
The neutral current mode can determine the core temperature precisely
even in the presence of the MSW effect:
assuming 10\% systematic errors, the high sensitivity of the
$^8$B neutrino flux
($\sim \Tc^{18}$)
allows a determination of $\Tc$ at the 0.5\% level.

\vspace{4.0ex}
\noindent
VII. SUMMARY
\vspace{2.0ex}

Existing Homestake, Kamiokande, and GALLEX experiments  strongly
disfavor astrophysical solutions invoking a cooler Sun.
For the matter-enhanced neutrino oscillations, the data constrain
the parameter space to two small regions, one in the non-adiabatic
region (which is preferred by the data)
and one in the large-mixing region.   The fit for oscillations
into a sterile neutrino allows only non-adiabatic oscillations at
90\% C.L.  \, Allowing a
non-standard core temperature along with MSW oscillations we find
that the data constrain the core temperature
at the 5\% level, yielding values consistent with the SSM.
The warmer Sun ($\Tc$=1.05) allows the parameter space predicted
by the SO(10) GUT with an intermediate-breaking scale
(for which the $\nutau$ may be cosmologically relevant),
while the cooler Sun stretches
the allowed parameters into a region predicted by simple
supersymmetric
SO(10) GUTs.  Superstring-inspired models are consistent with all
solutions.
Predictions are made for future solar neutrino detectors.

We thank John Bahcall and Marc Pinsonneault for providing us with
the solar model and neutrino fluxes in computer-readable form.
It is a pleasure to thank Eugene Beier and
Marc Pinsonneault for useful discussions.  One of us (N.H.) is
indebted to Ed Frank for invaluable advice in computations.
This work was supported by Department of Energy Contract
DE-AC02-76-ERO-3071 and DE-FG02-88ER40479.

\newpage

\parskip 2ex
\setlength{\parindent}{-1em}
{\large Table Captions}

Table~I.~~The $\Tc$ fits for various combinations of the Homestake (Cl),
Kamiokande II+III (Kam), and GALLEX results.  Listed are the best fit
value of
$\Tc$ with 90\% confidence level (C.L.)
error, $\chi^2$ values, and C.L. of excluding the fits.

Table~II.~~The best fit of the Homestake, Kamiokande II+III, and GALLEX
results.  In the $\Tc=1$ (SSM) column, shown are the best fit of
$\dm2$ and
$\s22t$ with the $\chi^2$ value for each of the allowed MSW regions.
  The $\Tc$ = free column is for the
  three parameter fit ($\dm2$, $\s22t$, and $\Tc$).
The $\Tc$ errors are at 90\% C.L.

Table~III.~~The best fit of the Homestake, Kamiokande II+III, and GALLEX
results for sterile-neutrino oscillations.
In the $\Tc=1$ (SSM) column, shown are the best fit of
$\dm2$ and
$\s22t$ with the $\chi^2$ value for each of the allowed MSW regions.
  The $\Tc$ = free column is for the
  three parameter fit ($\dm2$, $\s22t$, and $\Tc$).
The $\Tc$ errors are at 90\% C.L.

Table~IV.~~Predicted rates for future solar neutrino detectors,
relative to the SSM expectations.
The rates are listed for each of the allowed regions obtained from the
best fit (Fig.~6\,(a)).

\newpage
\parskip 2ex

\setlength{\parindent}{-1em}
{\large Figure Captions}

Figure~1.~~The approximate
$\Tc$ dependence of the neutrino counting rates (relative to
the SSM) for the Cl, Ga, and Kamiokande experiments, according to the
power laws (Eqn.~(1)).  $\Tc$ is relative to
the SSM value ($\Tc=1=15.67\times 10^6$ K = 1.35 keV).

Figure~2.~~The $\nue$ survival  probability contours (solid lines)
 for Cl experiments and the allowed region obtained from the Homestake
result (90\% C.L., shaded region).
The contours are for survival
probabilities of 0.1, 0.2,$\cdots$, 0.9, starting with the innermost
solid line.
The calculation of the allowed region
includes the experimental errors, the detector cross section
uncertainties, and the SSM flux errors.
The allowed region slightly deviates
from the iso-probability contours because of the difference in the
$^7$Be and $^8$B flux uncertainties.

Figure~3\,(a).~~The $\nue$ survival probability contours
(solid lines) for
Kamiokande experiments and the allowed region for the Kamiokande II and
III (220 days) result (90\% C.L., shaded region).
This is for flavor oscillations into $\numu$ or $\nutau$.
The contours are for effective survival
probabilities of 0.2, 0.3,$\cdots$, 0.9, which include the effects of
neutral current scattering. (There is no 0.1 contour.)
The calculation
includes the energy threshold, the energy resolution, and the trigger
efficiency.  The allowed region includes the SSM  uncertainties of
the $^8$B flux.

Figure~3\,(b).~~Same as Fig.~3\,(a), except that it is for
oscillations into a sterile neutrino.  The
contours are for survival probabilities of 0.1, 0.2,$\cdots$, 0.9.
Compared to flavor oscillations, lack of a neutral current contribution
increases the $\nue$ survival probability required by the data
and therefore pushes the allowed region outward of the triangle.

Figure~4.~~The $\nue$ survival probability contours (solid lines)
for Ga experiments and the 90\%-C.L. allowed region for the GALLEX result.
The calculation of the allowed region includes the experimental errors,
the detector cross section uncertainties, and the SSM flux errors.
The contours are for survival probabilities of 0.1, 0.2,$\cdots$, 0.9.

Figure~5.~~Same as Fig.~4, except the allowed region is for SAGE.

Figure~6\,(a).~~The allowed region for Homestake (dotted region),
Kamiokande II+III
(solid line), and GALLEX (dashed line) results.  The shaded region is
the combined fit of the three experiments (90\% C.L.).

Figure~6\,(b).~~The allowed region of combined Homestake,
Kamiokande II+III,  and GALLEX results at 68\% (shaded region), 90\%
(dashed lines), and 95\% (dotted lines) C.L.

Figure~6\,(c).~~The allowed region of the combined results {\em with} (dotted
line) and {\em without} (shaded regions) theoretical uncertainties from
the SSM
and the detector cross sections.

Figure~7.~~The combined fit of the Kamiokande II+III and Homestake results
(90\% C.L.).

Figure~8.~~The combined fit of the Kamiokande II+III and GALLEX results
(90\% C.L.).

Figure~9.~~The combined fit of the Homestake and GALLEX results
(90\% C.L.).

Figure~10.~~The fit to oscillations to sterile neutrinos at 68\% (shaded),
90\% (dashed line) and 95\% (dotted line) C.L.
The data are the combined result of the Homestake, Kamiokande II+III,
and GALLEX.

Figure~11.~~The $\chi^2$ plot as a function of $\Tc$ in the three
parameter fit ($\Tc, \dm2$, and $\s22t$) of the Homestake, Kamiokande,
and GALLEX.  For each $\Tc$, the $\chi^2$ is minimized with respect to
$\dm2$ and $\s22t$.  The data determine $\Tc$ within
$0.98 \leq \Tc \leq 1.06$ (90\% C.L.), which is consistent with the
SSM value.

Figure~12.~~The allowed MSW region of the combined result of Homestake,
Kamiokande II+III, and GALLEX, using $\Tc$ as a free parameter.  As a result
of allowing $\Tc$ to change, the 90\% C.L. region widens.  Also shown are
$\dm2$, $\s22t$ predicted by SUSY SO(10) GUTs (shaded), intermediate-scale
SO(10) GUTs (thick line), and string-inspired SUSY models with
nonrenormalizable operators (shaded line).  In each model the predictions
for  $\dm2$ are not robust and easy to change.  In the string-inspired
model, $\s22t$ is also changeable.

Figure~13\,(a).~~The allowed MSW region of the combined result of Homestake,
Kamiokande II+III, and GALLEX when $\Tc$ is fixed at 1.05 (a warmer Sun).
As a result of the high $\Tc$, the two regions of the SSM fit
(Fig.~6\,(a))
merge into one, allowing the parameters predicted by SO(10) GUTs with
intermediate-breaking scales.

Figure~13\,(b).~~The allowed MSW region of the combined result of Homestake,
Kamiokande II+III, and GALLEX when $\Tc$ is fixed at 0.95 (a cooler Sun).
As a result of the low $\Tc$, the allowed region is pushed outward of the
MSW region, allowing the parameters predicted by supersymmetric SO(10)
GUTs.

Figure~14~~The $\nue$ survival probabilities as a function of energy for
the two regions obtained by the best fit (Fig.~6\,(a)).
The solid line is for the non-adiabatic region
and the dashed line is for the large-mixing region.

Figure~15.~~The predicted spectral shape of charged-current events at
SNO.  The solid line is for the allowed parameter space in the non-adiabatic
region, and the dashed line is for the large-mixing region.  The errors
are equivalent to
a 2-year operation (6000 events).  The distortion of the
spectrum in the non-adiabatic branch will confirm the MSW
effect and differentiate the two allowed regions (Fig.~6\,(a)).

\newpage

\renewcommand{\thetable}{\Roman{table}}
\begin{table}[p]
\centering
\caption{$\Tc$ Fit for Kamiokande II+III, Homestake, and GALLEX}
\begin{tabular}{ l r r r }
\hline \hline 
                        & $\Tc   \pm   \Delta \Tc$  & $\chi^2$& C.L.   \\
\hline 
 Kam                    & 0.961 $\pm$ 0.010         & 0       & ---    \\
 Cl                     & 0.901 $\pm$ 0.015         & 0       & ---    \\
 GALLEX                 & 0.860 $\pm$ 0.042         & 1.31    & ---    \\
 Kam+Cl                 & 0.921 $\pm$ 0.013         & 13.78   &$>$99.99\\
 Kam+GALLEX             & 0.960 $\pm$ 0.010         & 2.64    & 89.26 \\
 Cl+GALLEX              & 0.902 $\pm$ 0.013         & 1.49    & 77.56 \\
 Kam+Cl+GALLEX          & 0.920 $\pm$ 0.012         & 15.46   & 99.99 \\
\hline \hline 
\end{tabular}
\end{table}
\newpage
\begin{table}[p]
\centering
\caption{ Best Fit of Kamiokande II+III, Homestake, and GALLEX}
\begin{tabular}{l c c c c }
\hline \hline 
         &\multicolumn{2}{c}{$\Tc=1$} &\multicolumn{2}{c}{$\Tc$=free}\\
         & Non-adiabatic & Large-mixing & Non-adiabatic & Large-mixing\\
\hline 
 $\s22t$ & 6.7$\times10^{-3}$ & 0.73 & 7.7$\times 10^{-3}$ & 0.31  \\
 $\dm2$(eV$^2$) & 6.1$\times10^{-6}$ & 1.1$\times 10^{-5}$
                       & 8.9$\times10^{-6}$ & 1.1$\times10^{-5}$ \\
 $\Tc$   & 1$\pm$0.0057 & 1$\pm$0.0057 & 1.03$^{+0.03}_{-0.05}$
                                               & 1.05$^{+0.01}_{-0.07}$  \\
 $\chi^2$&  0.56    & 3.52       & 0.          & 0.           \\
\hline \hline 
\end{tabular}
\end{table}
\begin{table}[p]
\centering
\caption{ Best Fit for Oscillations into Sterile Neutrino}
\begin{tabular}{l c c c c }
\hline \hline 
         &\multicolumn{2}{c}{$\Tc=1$} &\multicolumn{2}{c}{$\Tc$=free}\\
         & Non-adiabatic & Large-mixing & Non-adiabatic & Large-mixing\\
\hline 
 $\s22t$ & 9.7$\times10^{-3}$ & 0.83 & 7.5$\times 10^{-3}$ & 0.15  \\
 $\dm2$(eV$^2$) & 3.8$\times10^{-6}$ & 4.8$\times 10^{-6}$
                       & 4.0$\times10^{-6}$ & 3.5$\times10^{-6}$ \\
 $\Tc$   & 1$\pm$0.0057 & 1$\pm$0.0057  & 0.99$^{+0.16}_{-0.02}$ &
                                           1.15$^{+0.005}_{-0.18}$ \\
 $\chi^2$& 3.17         &  8.95         & 3.13        & 4.91         \\
\hline \hline 
\end{tabular}
\end{table}
\newpage
\begin{table}[p]
\centering
\caption{ Predicted Rates for Future Detectors}
\begin{tabular}{l  c c }
\hline \hline 
         &\multicolumn{2}{c}{Rate / SSM}  \\
                                & Non-adiabatic & Large-mixing     \\
\hline 
SNO (Charged Current)           &  0.15 - 0.5   &   0.15 - 0.3     \\
SNO ($\nu-e$ scattering)        &  0.3  - 0.6   &   0.3  - 0.45    \\
Super-Kamiokande                &  0.3  - 0.6   &   0.3  - 0.45    \\
Borexino ($^7$Be$\nu-e$)        &  0.15 - 0.6   &   0.3  - 0.6     \\
\hline 
\hline 
\end{tabular}
\end{table}

\end{document}